\documentclass[12pt]{article}

\usepackage{graphics,epsfig}
\begin{document}
\title{
Minority Opinion Spreading \\in \\Random Geometry
}
\author{Serge Galam
\\Laboratoire des Milieux D\'esordonn\'es et H\'et\'erog\`enes, \\
Tour 13, Case 86, 4 place Jussieu, 75252 Paris Cedex 05, }

\date{(galam@ccr.jussieu.fr)}
\maketitle

\begin{abstract}

The dynamics of spreading of the minority opinion in public debates (a reform 
proposal, a behavior change, a military retaliation) is studied using 
a diffusion reaction model.  People move 
by discrete step on a landscape of random geometry shaped by social life 
(offices, houses, bars, and restaurants). A perfect world is 
considered with no advantage to the minority. A one person-one argument 
principle is applied to determine locally individual mind changes. In case 
of equality, a collective doubt is evoked which in turn favors the Status Quo.
Starting from a large in favor of the proposal initial majority, repeated random size
local
discussions are found to drive the majority reversal along the minority hostile view.
Total opinion refusal is completed within few days. Recent 
national collective issues are revisited. The model may apply to rumor and fear
propagation.

\end{abstract}

\newpage

  All over the world and more specifically in democratic countries public opinion
 seems to be rather conservative while facing a nationwide issue open to a public 
debate like for instance a reform proposal or a behavior change [1, 2, 3]. Even when 
the changes at stake are known to be desperately needed (medical evidences, danger 
of death, administrative inefficiencies) an initial hostile minority appears to be 
almost always able to turn the majority along its refusal position. 

A symptomatic illustration of such a paradoxical social refusal was the year 2000 
generalized failure of the French government to reform the academic system, the 
taxes collect and the agriculture system of economical help [3]. Another example 
is the Irish No to the Nice European treaty that came as a surprise to the Irish 
people itself [4]. Along with this reality some people could be tempted 
to consider reforms possible only using social violence or authoritarian 
top leadership decisions. It thus arises he fundamental question whether 
or not a reform can be decided democratically at least in principle. 
  
To understand the reason of such a social inertia, most research has concentrated 
on analyzing the complicated psycho-sociological mechanisms involved in the process 
of opinion forming. In particular focusing on those by which a huge majority of people 
gives up to an initial minority view [1, 2]. The main feature being that the prospect 
to loose definite advantages is much more energizing than the hypothetical gain of a 
reform. Such an approach is certainly realistic in view of the very active nature of 
minorities involved in a large spectrum of situations.
  
However in this letter we claim that in addition to the more aggressiveness and 
persuasive power of a threatened or very motivated minority there exists some basic 
and natural mechanism inherent to free public debate which makes the initial
hostile minority to a full spreading over.
  
To ground our claim we present an extremely simple model to opinion forming using 
some concepts and techniques from the physics of disorder [5, 6, 7, 8]. A
diffusion reaction model is implemented on a landscape of random geometry. 
It does not aimed at an exact description of reality. 
But rather, by doing some crude 
approximations, it focuses on enlightening
an essential feature of an otherwise very complex and multiple phenomena. 
In particular the holding of free public debate is shown to lead almost systematically
to the total spreading of an initial hostile minority view within the initial proposal
in favor huge majority.
The associated dynamics of extreme public polarization at the 
advantage of the initial minority is found to result from the existence 
of asymmetric unstable thresholds [9, 10] that are produced by the 
random occurrence of temporary local doubts. Some recent nation wide issues 
with respect to European construction are thus revisited [4, 11]. The application 
to the phenomena of rumor and fear propagation is discussed [12].

We start from a population with $N$ individuals, which have to decide whether or not 
to accept a reform proposal. At time $t$ prior to the discussion the proposal has 
a support by $N_+(t)$ individuals leaving $N_-(t)$  persons against it.  Each person 
is supposed to have an opinion making $N_+(t)+N_-(t)=N$.  Associated individual 
probabilities to be 
in favor or against the proposal are thus,

\begin{equation}
P_{\pm}(t)\equiv \frac{N_{\pm}(t)}{N}  ,
\end{equation} 
with,
\begin{equation}
P_+(t)+P_-(t)=1 .
\end{equation}

From this initial configuration, people start discussing the project. However 
they don't meet all the time and all together at once. Gatherings are shaped 
by the geometry of social life within physical spaces like offices, houses, 
bars and restaurants. This geometry determines the number of people, which meet 
at a given place. Usually it is of the order of just a few. Groups may be larger 
but it is much rare.  Accordingly a given social life yields a random local
geometry landscape characterized by a probability 
distribution for gathering sizes $\{a_i\}$ which
satisfy the constraint,

\begin{equation}
\sum_{i=1}^L a_i=1 ,
\end{equation}

where $i=1, 2, ..., L$  stands 
for respective sizes $1, ..., L$ with $L$ being the larger group. 
While discussions 
can occur by chance, most are monitored through regular time break meetings 
like lunch, dinner, happy hour and late drink. At each encounter people meet 
with different people (friends, colleagues and acquaintances) at different areas 
of their social space. It extends from a building, a neighborhood, a town or another
 state. Thus, people gatherings occur in sequences in time, each one allowing a new 
local discussion. There, people may change their mind with respect to 
the reform proposal. During these meetings all individuals are assumed to be involved 
in one group gathering. It means a given person is, on average, taking part to a 
group of size $i$  with probability $a_i$. The existence of one-person groups makes 
this assumption realistic. Each new cycle of multi-size discussions is marked by a time 
increment $+1$. 

To emphasize the bare mechanism at work into the refusal dynamics which arises 
from local interactions we consider a perfect world.  No advantage is given to the 
minority with neither lobbying nor organized strategy. Moreover an identical 
individual persuasive power is assumed for both sides. A one person - one argument 
principle is used to implement the psychological process of collective mind update. 
On this basis a local majority argument determines the outcome of the discussion. 
People align along the local initial majority view. 

For instance a group of five 
persons with at start three in favor of the reform and two against it ends up with 
five people in favor of the reform. On the reverse two initial reform supporters 
leads to five persons against it.
However in case of an even group this rule of one person - one argument leaves 
the local possibility of a temporary absence of a collective majority. The group 
is then at a tie within a non-decisional state. It doubts. 

There we evoke a physical 
principle called the ``inertia principle" which states that to put on motion a system, 
it is necessary to apply on it a force at least infinitesimally larger than the 
friction which holds on it to keep the system at rest. This principle can be put 
in parallel to the fundamental psychological asymmetry that exists between what 
is known and what is hypothetical. Therefore, to go along what is unknown, even if 
this unknown is supposed to be better, a local majority of at least one voice is 
necessary. 

In terms of our model, at a tie the group does not move and thus decides 
not to move, i. e., the full group turns against the reform proposal to preserve the 
existing situation. For instance a group of six persons with initially three in favor 
of the reform and three against it yields six persons against the reform. It is worth 
to stress that this is not an advantage given to the minority in terms of being more 
convincing. It is a collective outcome that results from a state of doubt. For instance
 dealing with a military retaliation people will avoid action unless there exist clear 
evidences for it. Accordingly having  $P_{\pm}(t)$ at time $t$ yields,

\begin{equation}
P_{+}(t+1)=\sum_{k=1}^L a_k \sum_{j=N[\frac{k}{2}+1]}^k C_j^k P_+(t)^j P_-(t)^{(k-j)} ,
\end{equation}
at time $(t+1)$ where $C_j^k\equiv  \frac{k!}{(k-j)! j!}$ and 
$N[\frac{k}{2}+1]\equiv IntegerPart \ of \ (\frac{k}{2}+1)$. Simultaneously,
\begin{equation}
P_{-}(t+1)=
\sum_{k=1}^L a_k \sum_{j=N[\frac{k}{2}]}^k C_j^k P_-(t)^j P_+(t)^{(k-j)} .
\end{equation}

In the course of time, the same people will meet again and again
randomly in the same cluster configuration. At each new encounter they
discuss locally the issue at stake and change their mind according to
above majority rule. To follow the time evolution of the reform 
support Eq. (4) is iterated until a 
stable value is reached. A monotonic flow is obtained towards either one of 
two stable fixed points $P_{+N}=0$ and $P_{+Y}=1$. The flow and its direction are 
produced by an 
unstable fixed point $P_{+F}$ located in between $P_{+N}$ and $P_{+Y}$. Its value 
depends on both the $\{a_i\}$ and $L$.  We denote it the Faith point. For 
$P_+(t)<P_{+F}$  it exists a number $n$  such that  $P_+(t+n)=P_{+N}=0$ 
while for $P_+(t)>P_{+F}$  it is another number $m$  which yields  $P_+(t+m)=P_{+Y}=0$.
Both $n$ and $m$  measure the 
required time at reaching a stable and final opinion. It is either a ``Big Yes" 
to the reform at $P_{+Y}=1$ or a ``Big No" at $P_{+N}=0$. Their respective values 
depend on the $\{a_i\}$, 
$L$  and the initial value $P_+(t)$ .

Repeated successive local
discussions thus drive the whole population to a full polarization 
with a ``Big Yes" to 
the reform project at $P_{+Y}=1$, or a ``Big No" 
at $P_{+N}=0$. 
Accordingly, public opinion is not volatile. It stabilizes rather quickly 
($n$ and $m$ are usually small numbers) to a 
clear stand.

Figure 1 shows the variation of $P_{+}(t+1)$ as function of $P_{+}(t)$ for two 
particular sets of the $\{a_i\}$. First one is $a_1=a_2=a_3=a_4=0.2$ and $a_5=a_6=0.1$ 
where $L=6$. There
$P_{+F}=0.74$ which puts the required initial support to the reform success
at a very high value of more than $74\%$.
Simultaneously an initial minority above $26\%$ is enough to produce a final total 
refusal. The second set is $a_1=0$, $a_2=0.1$ and $a_3=0.9$ 
with $L=3$ and $P_{+F}=0.56$. There the situation is much milder but 
also unrealistic since always pair discussions are much more numerous than just $10\%$.

\begin{figure}
\epsfxsize=\columnwidth
\begin{center}
\centerline{\epsfbox{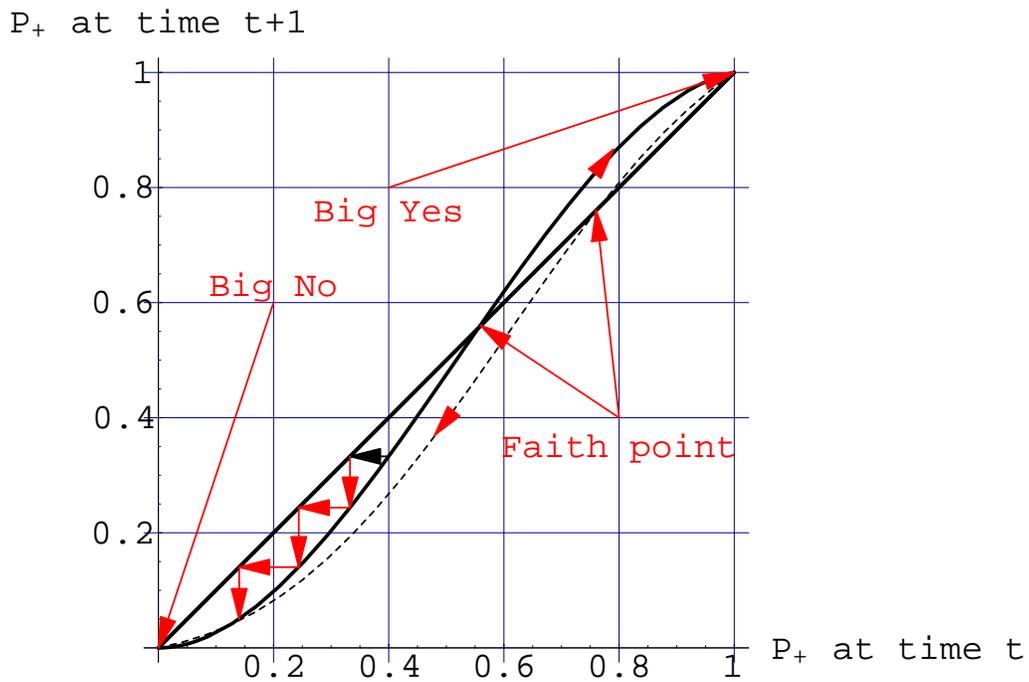}}
\caption{Variation of $P_{+}(t+1)$ as function of $P_{+}(t)$. The dashed line is for 
the set $a_1=a_2=a_3=a_4=0.2$, $a_5=a_6=0.1$, 
$L=6$ and $P_{+F}=0.74$. The plain line is for the set $a_1=0$, $a_2=0.1$ and $a_3=0.9$ 
with $L=3$ and $P_{+F}=0.56$. 
Arrows show the direction of the flow.}

\end{center}
\end{figure}

To make a quantitative illustration of the dynamics refusal let us consider 
above first setting with an initial $P_{+}(t)=0.70$  at time  $t$. The associated 
series in time is  $P_{+}(t+1)=0.68$, 
$P_{+}(t+2)=0.66$, $P_{+}(t+3)=0.63$, $P_{+}(t+4)=0.58$,
$P_{+}(t+5)=0.51$, $P_{+}(t+6)=0.41$, $P_{+}(t+7)=0.27$, $P_{+}(t+8)=0.14$,
$P_{+}(t+9)=0.05$, $P_{+}(t+10)=0.01$ and eventually $P_{+}(t+11)=0.00$. Eleven cycles 
of discussion 
make all $70\%$ of reform supporters to turn against it by merging with the initial 
$30\%$ of reform opponents. On a basis of one discussion a day on average, less than 
two weeks is enough to a total crystallization of the No against the reform proposal. 
Moreover a majority against the reform is obtained already within six days 
(see Fig. (2)).

Changing a bit the parameters with $a_1=0.2$, $a_2=0.3$, $a_3=0.2$, 
$a_4=0.2$, $a_5=0.1$ and $a_6=0$ gives $P_{+F}=0.85$, a higher value, which makes any 
reform proposal quite impossible. How a realistic reform project could start 
with already more than $85\%$ support in the population? Starting still from  
$P_{+}(t)=0.70$ yields successively $P_{+}(t+1)=0.66$, 
$P_{+}(t+2)=0.60$, $P_{+}(t+3)=0.52$, $P_{+}(t+4)=0.41$,
$P_{+}(t+5)=0.28$, $P_{+}(t+6)=0.15$, $P_{+}(t+7)=0.05$, $P_{+}(t+8)=0.01$ before
$P_{+}(t+9)=0.00$. The number of local 
meetings has shrieked from $11$ to $9$. Within ten days the whole population stands 
against the reform proposal. The initial $30\%$ of opponents grow to more than fifty 
percent in less than four days (see Fig. (2)).

\begin{figure}
\begin{center}
\centerline{\epsfbox{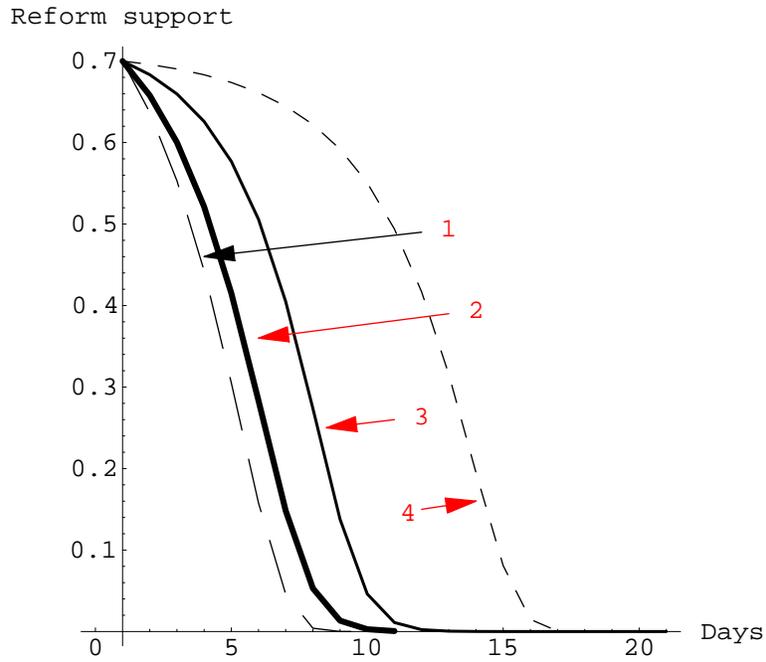}}
\caption{Variation $P_{+}(t)$ as function of successive days with $L=6$. 
The initial value  at $t=1$ is $P_{+}(1)=0.70$. 
Long dashed line (1): $a_1=0$, $a_2=\frac{1}{2}$, $a_3=\frac{1}{2}$, $a_4=a_5=a_6=0$
with $P_{+F}=1$.  
Heavy thick line (2): $a_1=0.2$, $a_2=0.3$, $a_3=0.2$, $a_4=0.2$, $a_5=0.1$ 
and $a_6=0$ with $P_{+F}=0.85$. 
Other line (3): $a_1=a_2=a_3=a_4=0.2$, $a_5=a_6=0.1$. There $P_{+F}=0.74$. 
Dashed line (4): $a_1=0$, $a_2=0.3$, $a_3=0.7$, $a_4=a_5=a_6=0$ with $P_{+F}=0.71$.}
\end{center}
\end{figure}    

It is the existence of an unstable fixed point between the 
two stable ones which produces the whole polarization dynamics. The 
stable ones are constant and independent of the {$a_i$} but 
the unstable one varies with both, sizes and the {$a_i$} distribution.
To single out the specific contribution of each gathering size to the aggregation 
effect we now determine the associated unstable fixed point for groups from two 
to six. Values are shown in Table 1. The flow landscape is identical for all odd 
sizes with an unstable fixed point at $\frac{1}{2}$. On the opposite for even sizes the 
unstable fixed point starts at one for size two and decreases to $0.65$ at size six, 
via $0.77$ at size four.

\begin{table}
\caption{\sf Values of the various fixed points for each group size from two to six.
$STP\equiv$ Stable fixed point and $UTP\equiv$ Unstable fixed point}
\label{tbl}
\begin{center}
\begin{tabular}{|l|l|l|l|}
\hline
Group  &SFP  &UFP   &SFP \\ 
Size & Total No $P_{+N}$&Faith Point $P_{+F}$& Total Yes $P_{+Y}$\\ 
\hline
2 & 0&1 &  none\\ 
\hline
3 & 0&$\frac{1}{2}$ &  1\\ 
\hline
4 & 0&$\frac{1+\sqrt{13}}{6}\approx 0.77$ &  1\\ 
\hline
5& 0&$\frac{1}{2}$ & 1 \\ 
\hline
6 & 0&$\approx 0.65$& 1\\ 
\hline
\end{tabular}
\end{center}
\end{table}


To illustrate the interplay dynamics between even and odd 
sizes, let us look at more details in the hypothetical case of discussion groups 
restricted to only two and three persons. Putting $a_1=a_4=..=a_L=0$,
Eq. (4) reduces to,

\begin{eqnarray}
 P_{+}(t+1) & = & a_2 P_+(t)^2 \\
\nonumber& & 
+(1-a_2)\{P_+(t)^3 +3 P_+(t)^2 (1-P_+(t))\},
\end{eqnarray}
whose stable fixed points are still $0$ and $1$ with the unstable one lacated at,

\begin{equation}
P_{+F}=\frac{1}{2(1-a_2)}.
\end{equation}

For $a_2=0$ (only three size groups) we recover  $P_{+F}=\frac{1}{2}$
while it gets to $P_{+F}=1$ already at $a_2=\frac{1}{2}$. 
It shows the existence of pair discussion has a drastic effect on creating doubt 
that in turn produces a massive refusal spreading. Few days are now enough to 
get a total reform rejection.

Keeping an initial  $P_{+}(t)=0.70$ the time series become $P_{+}(t+1)=0.64$, 
$P_{+}(t+2)=0.55$, $P_{+}(t+3)=0.44$, $P_{+}(t+4)=0.30$,
$P_{+}(t+5)=0.16$, $P_{+}(t+6)=0.05$ before $P_{+}(t+7)=0.00$. 
The reform supporters falling down is extremely sharp as shown in Fig. (2). 
Within a bit more than two days a majority of the people is already standing 
against the reform that yet started with a seventy- percent support. Seven days 
latter the proposal is completely out of any reach with not one single supporter. 
Considering instead only thirty percent of pair discussion groups, the falling is 
weakened but yet within sixteen days we have  $P_{+}(t+16)=P_N0.00$ (see Fig. 2). 

Clearly an infinite number of combinations of the $\{a_i\}$ is possible. 
However the existence of these temporary local doubts which ultimately 
produces a strong polarization towards social refusal is always preserved. 
At this stage it is worth to stress that in real life situations not every 
person is open to a mind change. Some fractions of the population will keep 
on their opinion whatever happens. Including this effect in the model will not 
change qualitatively the results. It will make the polarization process not total 
with the two stable fixed points shifted towards respectively larger and smaller 
values than zero and one. 

To give some real life illustrations of our model, we can cite events related to 
the European Union which all came as a surprise. From the beginning of its 
construction there have been never a large public debate in most of the involved 
countries. The whole process came trough government decisions tough most people 
always have seemed to agree on this construction. At the same time European 
opponents have been systematically urging for public debates. Such a demand 
sounds like absurd knowing a majority of people favor the European union. 
But anyhow most European governments have been reluctant to held referendum 
on the issue. 

At odd, several years ago French president Mitterand decides to run a referendum 
to accept the Maastricht agreement [11]. While a large success of the Yes was
given 
for granted it indeed made it just a bit beyond the required fifty percent. 
The more people were discussing, the less support there was for the proposal. 
It is even possible to conjuncture that an additional two weeks extension of 
the public debate would have make the No to win. The very recent Irish No [4] 
which came as a blow to all analysts may obey the same logic. The difference 
with the French case was certainly the weaker initial support. Of course addition 
political reasons to the No were also active.

To conclude, even tough our model is clearly very crude it does demonstrate the 
inherent polarization effect associated to the holding of democratic debates 
towards social immobility. Moreover this process was shown to be de facto 
anti-democratic since even in a perfect world it makes an initial minority 
refusal to almost systematically spread over in convincing very quickly the 
whole population. The existence of natural random local temporary doubts is 
instrumental in this phenomenon driven by the geometry of social life. As on how 
to remedy this reversal phenomenon, the first hint is to avoid the diffusion of 
the No via these localized temporary doubts. Direct and immediate votes would 
address this task preserving the democratic expression of an unbiased initial 
majority. But more thoughts and studies should be performed before getting to 
a clear proposal scheme. The model may generalize to a large spectrum of social, 
economical and political phenomena that involve propagation effects. In 
particular it could shed a new light on both processes of fear propagation 
and rumors spreading.


\begin{thebibliography}{12}

\bibitem {1} R. D. Friedman and M. Friedman, The Tyranny of the Status Quo, 
Harcourt Brace Company (1984) 
\bibitem {2} S. Moscovici, Silent majorities and loud minorities, 
Communication Yearbook, 14. J. A. Anderson. Sage Publications Inc. (1990)
\bibitem {3} S. Galam Les r/'eformes sont-elles impossibles ?, 
Le Monde/ 28 mars/ 18-19 (2000)
\bibitem {4} http://news.bbc.co.uk/hi/english/world/europe/ 
Prodi tackles Irish sceptics (2001)
\bibitem {5} S. Galam, B. Chopard, A. Masselot and M. Droz, 
Eur. Phys. J. B 4, 529-531 (1998) 
\bibitem {6} M. Barth/'el/'emy and L. A. N. Amaral, 
Phys. Rev. Lett. 82, 3180-3183 (1999)
\bibitem {7} S. Solomon, G. Weisbuch, L. de Arcangelis, N. Jan and D. Stauffer,
Physica A277 (1-2), 239-247 (2000) 
\bibitem {8} Gerard Weisbuch, Guillaume Deffuant, Frederic Amblard, Jean Pierre Nadal,
cond-mat/0111494
\bibitem {9} T. C. Schelling Micromotives and Macrobehavior, New York,
Norton and Co. (1978)
\bibitem {10} M. Granovetter, American 
Journal of Sociology 83, 1420-1443 (1978)
\bibitem {11} M. Franklin, M. Marsh and L. McLaren Uncorking the Bottle: Popular 
Opposition to European Unification in the Wake of Mastricht, Annual meeting 
of the Midwest Political Science Association (1993)
\bibitem {12} R. Pastor-Satorras and A. Vespignani, 
Phys. Rev. Lett. 86, 3200-3183 (2001)

\end{thebibliography}
\end{document}